# Laser Induced Damage Studies in Borosilicate Glass Using nanosecond and sub nanosecond pulses

Vinay Rastogi[a,b], S. Chaurasia[a, *], D. S. Munda[a]

[a]High Pressure and Synchrotron Radiation Physics Division, Physics Group, Bhabha Atomic Research Centre, Trombay Mumbai-400085, India
[b]Homi Bhabha National Institute, Mumbai-400094, India
[*]pgshivanand@gmail.com

**Abstract**

The damage mechanism induced by laser pulse of different duration in borosilicate glass widely used for making confinement geometry targets which are important for laser driven shock multiplication and elongation of pressure pulse, is studied. We measured the front and rear surface damage threshold of borosilicate glass and their dependency on laser parameters. In this paper, we also study the thermal effects on the damage diameters, generated at the time of plasma formation. These induced damage width, geometries and microstructure changes are measured and analyzed with optical microscope, scanning electron microscope and Raman spectroscopy. The results show that at low energies symmetrical damages are found and these damage width increases nonlinearly with laser intensity. The emitted optical spectrum during the process of breakdown is also investigated and is used for the characterization of emitted plasma such as plasma temperature and free electron density. Optical emission lines from Si I at 500 nm, Si II at 385nm and Si III at 455 nm are taken for the temperature calculations.

## 1. Introduction

The laser induced damage studies in optical glasses are important for various applications; such as making confinement geometry targets, damage morphology at interaction zone,laser micromachining, breakdown and excitation mechanism studies in optical components etc.[1]–[6]. Lasers with pulse duration from femtosecond to nanosecondinitiates inducing damage on the surface of optical component, ceramic, transparent materials through breakdown, sputtering or ablation processes[6]. The laser induced damage morphology by picosecond and nanosecond lasers has studied by many researchers. However a comparison of surface morphology changes induced by laser pulses has not been yet fully understand.

When an optical component is irradiated by a high power focused laser beam, a hot and dense plasma generated within a fraction of a nanosecond. This plasma generates a recoil pressure into the target, which may go up to the order of GPa and drives a compression wave into the target. These laser induced compression waves have so many applications such asstudy of materials behavior under extreme conditions, surface densification of porous and metallic materials, phase transition, spallation of ductile and brittle materials and so on[7]–[9].In comparison to conventional dynamic shock loading a high pressure and high strain can be achieved by laser shock loading for a nanosecond time scale.The material behavior under such dynamic shock loading at the molecular level can be examined with the help of time resolved Raman spectroscopy technique[10], [11].Thus, high power pulsed laser gives a new option for shock generation in spite of traditional projectile or explosive loading techniques.

The laser shock experiment can be classified into two types: Direct ablation and confined ablation. The direct ablation technique has problem of short pressure pulse which lead to the damping of compression wave due to averaging of rarefaction wave. Anderholm[12]was the first to introduce the concept of increasing the pressure by confining the plasma using a dielectric such as glass or water during its expansion. After that several studies has been done by using confinement geometry targets to investigate materialbehavior under extreme conditions.The main advantage of using a transparent layer of dielectric material over the target for laser produced plasma confinement is that pressure up to ten times of magnitude can be achieved easily than direct ablation at the same power density and the duration of shock waves may also be amplified by 2-3 times[13].To optimize such experiments using confinement geometry targets, a detailed characterization of laser irradiated optical glass are important[14], [15].

The main aim of the present work is to measure and understand the damage mechanisms in the plasma confining glass, used for making confinement geometry targets. In this paper, we have investigated the effect of laser pulse durations, influence of laser wavelength and the effect of thermal stresses on the glass sample. At the same time the plasma parameters are also estimated by analyzing the emitted optical spectrum in the breakdown process. The resultsof the present experiment will improve our understanding of the laser induced damage mechanisms in the dielectric materials and provide data for dielectric breakdown, which is the main limiting process in the confined geometry targets.

## 2. Experimental Method

2.1Laser Induced Damage Threshold (LIDT) Test

Laser induced damage threshold measurement and tests were performed on a borosilicate glass (BK7 glass). The schematic of the experimental setup for damage test is shown in figure 1. The damage threshold studies have been done for various laser parameters, such as its pulse duration (500 ps and 8 ns) and wavelengths (1064nm and 532nm).The frequency of 1064 nm (1$\omega$) of Nd: glass laser system was doubled by using KD$^*$P crystal. A beam reflected by beam splitter was incident on energy meter EM1 to monitor the shot to shot energy of the incident laser beam. The incident laser beam is focused on front and rear surface of glass sample using a short focal length lens L (f=5 cm). The effective,focused radius of the incident laser beam on the target is around 50 μm. Since glass transmission is an important parameter for the preparation of confinement geometry targets, an energy meter EM2 (Ophir Vega) is placed behind the glass sample to record the percentage transmitted laser radiation from the back side of the glass sample. The damage measurement was done by using three diagnostics; a photodiode, optical spectrometer and an optical CCD camera.Photodiode and optical spectrometerare used to collect the reflected laser and emitted spectrum from the sample respectively. When the damage starts the amplitude of reflected signal in the photodiode increases greatly. At the same time spectrometer recorded the emitted plasma spectrum from BK7 glass.

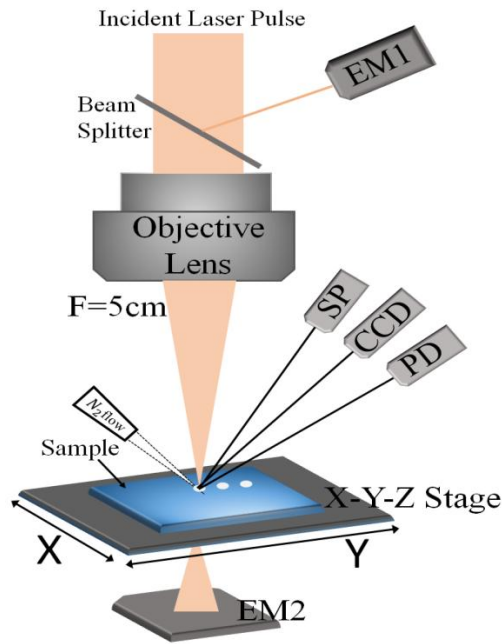

Fig. 1 Schematic of experimental setup; where EM1 and EM2-are energy metersused for the calibration of incident beam and transmitted laser pulse energy respectively; PD- photodiodeand CCD- charge coupled device camera are used for breakdown measurement; SP- optical spectrometer was used for recording the emitted optical spectrum

The background signal was recorded by the spectrometer and subtracted from the emission spectrum. This optical emission spectrum will give important information about the plasma parameters such as plasma temperature, electron density etc. The focal spot area was imaged by a CCD camera. At the time of damage; the generated damaged spot was recorded by the CCD camera connected to a personal computer. All these three diagnostics together will give a better measurement of damage threshold of the glass sample. The irreversible modifications in the glass sample were observed by an optical microscope. The entire damage test was performed on 1-on-1 procedure (single shot), to see the modification on the front and rear surface of the sample.To see the thermal effects generated in the plasma formation process due to high local temperature, we introduce a continuous flow of inert gas ($N_2$ gas) on the laser focused surface during the experiment.The heat generated at the plasma formation time is taken away from the surface of the sample by the flowing gas. Which reduces the effect of thermal stress on the surrounding area of the induced damage sites.

The damage modifications on the surface of the BK7 glass irradiated with pulse duration 500ps and 8 ns & with wavelength 1064 nm and 532 nm (2ω) for different laser intensity or energy were examined by SEM (Mini SEM SNE-3000M) with a resolution of 100 nm. The pulsed Raman spectrumof undamaged and damage regions had been taken by Raman spectrometer (Andor 550i with 3 cm$^{-1}$ resolution) with an ICCD camera at -30$^0$ temperatures. The excitation line used for Raman spectroscopy is 532 nm.

## 3. Results and Discussions

3.1Damage threshold and thermal effect studies of BK7under different pulse duration

The glass sample was irradiated with two different wavelengths (λ= 1064 nm and 532 nm) and two different laser pulse durations ($t_p$= 500 ps and 8 ns).The damage threshold for different cases are measured and given in table 1. Each data presented in the table 1 is an average over 10 shots.From the table 1 it is clear that the laser induced breakdown threshold energy is lower for 532 nm than 1064 nm. Which is due to the wavelength dependency of the damage threshold as described in several models[16], [17]. According to the pure multiphoton ionization model, with the increase in photon energy, the damage threshold exhibits an abrupt drop.Since the number of photons needed to bridge the band gap decreases, and the transition will take place at photon energies of$(\varepsilon_g - \Delta)/n$, where $\varepsilon_g$ and $\Delta$ are the band gap of the material and binding energy associated with an excitation respectively[2], [3], [18].Generally the band gap of materials is related with absorption coefficient (A), which is inversely proportional to wavelength. Hence laser induced breakdown threshold energy is lower for 532 nm than 1064 nm.From the table 1 it can also be seen that the damage threshold is lower for picosecond pulse in comparison to nanosecond pulse. Since the damage threshold energy depends on laser pulse duration ($t_p$) as $t_p^\alpha$where α varies from 0.3 to 0.6[2][19]. For our case the value of α is approximately equal to 0.3. The main physical processes involved at the present energies are also listed in table 1.

Table1. Threshold measurement for BK7 glass

| Laser | Pulse Duration $t_p$ (FWHM) | Wavelength $\lambda$ | Front surface Threshold (mJ) $E_{th}$ | Rear surface Threshold (mJ) $E_{th}$ | Main Physical process |
|---|---|---|---|---|---|
| Nanosecond | 8 ns | 1064 nm | 0.74 | 1.48 | Sputtering |
| Nanosecond | 8 ns | 532 nm | 0.36 | 0.59 | Sputtering |
| Picosecond | 500 ps | 1064 nm | 0.45 | 0.625 | Melting |
| Picosecond | 500 ps | 532 nm | 0.234 | 0.34 | Melting |

To understand the effect of laser irradiation on target, the damage morphologies are observed using an optical photo microscope and SEM. The Damaged area of 1064 nm is greater than the 532 nm. This difference in damage area is due to high dependence of surface carriers on material ionization band gap with laser wavelength[1], [4].Damage site images at constant pulse energy (4mJ), different pulse duration, wavelength and with inert gas flow are shown in figure2a-h. The damage craters have a spalled central part surrounded by micro cracks. These cracks formed due to a sharp increase in local temperature with increasing laser intensity. This heat propagates in 3- dimensions (due to Gaussian profile of laser pulse) through the material according to Fourier's law of heat conduction

$$\frac{dQ}{dt} = -k\nabla T \qquad (1)$$

Where k is known as thermal conductivity and $\nabla T$ is the temperature gradient. Fourier's law is empirical and describes the thermal diffusion of heat, analogous to Fick's first law of diffusion. The diffused heat in the lattice firstly reaches to the boiling point of the sample and producing high local pressure and temperature at the focal spot, leading to the distribution of additional stresses. Cracks are generated when these stresses exceed the maximum critical tensile or compressive strength. Due to the diffusion of this high local temperature in the surrounding of the sample, a circular ablated modification of glass occurred. The diameter of this circular deformation is decided by the intensity of the incident laser. A simple quantitative relation between ablation diameter (D), material dependent damage threshold fluence($\emptyset_{th}$), and peak fluence in the beam ($\emptyset_0$), is given by:

$$D^2 = 2\omega_0^2 ln\left(\frac{\emptyset_0}{\emptyset_{th}}\right) \qquad (2)$$

Where $\emptyset_0 = \frac{2I_p \times Area \times t_p}{\pi \omega_0^2}$, and $\omega_0^2 = \frac{1}{e^2}$, $I_p$ are Gaussian beam radius and laser pulse intensity respectively.

Also the pattern of these generated micro cracks changes with laser intensity, due to the increase in pressure or tensile stress in the surrounded area with the expansion of ablated center. The damage morphology for 532 nm is found to be same as of 1064 nm.

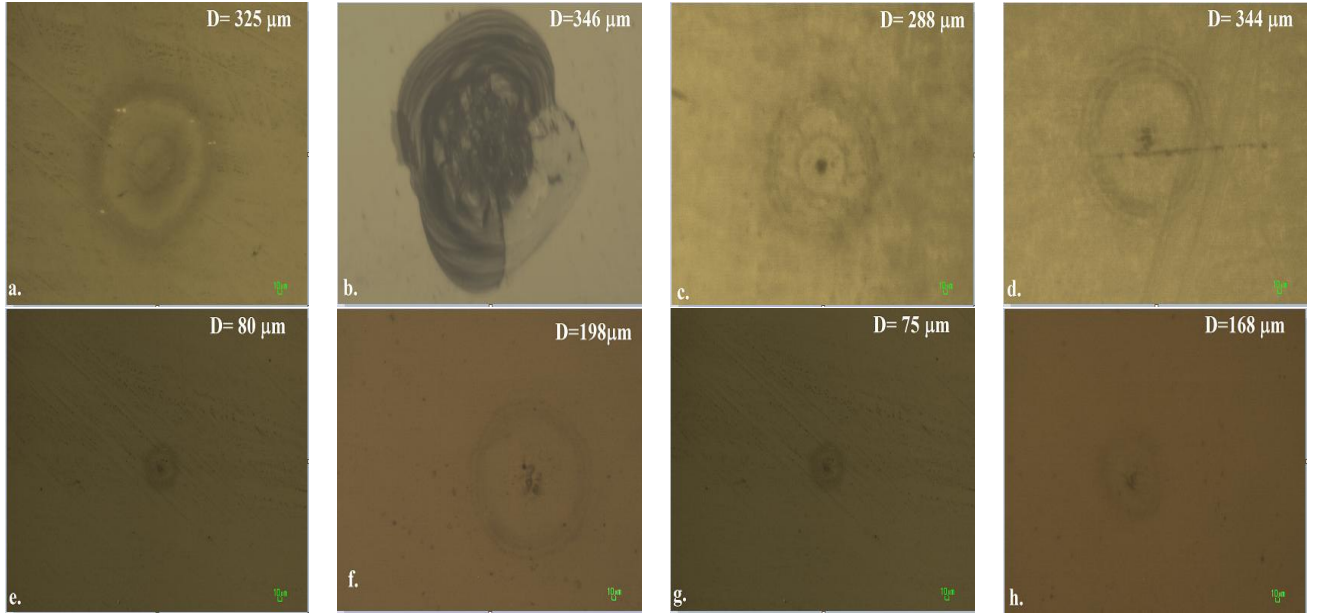

Fig. 2 Optical microscope images of the front surface of the sample at 4 mJ energy where;(a) λ= 1064nm, $t_p$= 500 ps without $N_2$ flow (b) λ= 1064nm, $t_p$= 8 ns without $N_2$ flow (c) λ= 532 nm, $t_p$= 500 ps without $N_2$ flow (d) λ= 532nm, $t_p$= 8ns without $N_2$ flow (e) λ= 532nm, $t_p$= 500ps with $N_2$ flow (f) λ= 1064nm, $t_p$= 8 ns with $N_2$ flow (g) λ= 532 nm, $t_p$= 500ps with $N_2$ flow (h) λ= 532nm, $t_p$= 8 ns with $N_2$ flow. Magnified 6x

To study the local thermal effect initiated by high local temperature and pressure on the damage threshold and morphology, we remove these thermal effects by continuous flow of inert gas on the surface of the sample. The images are shown in figure 2e-h. From the images it is clear that the local heat generated in the ablation process during the central crater formation is taken away by the inert gas. So the effect of thermal tensile stress is very much reduced and consequently damaged area decreases. In figure (2a-h) a comparison is done between the damage morphology for pulse durations $t_p$= 500 ps and $t_p$= 8 ns. More detail study is presented in section 3.3.

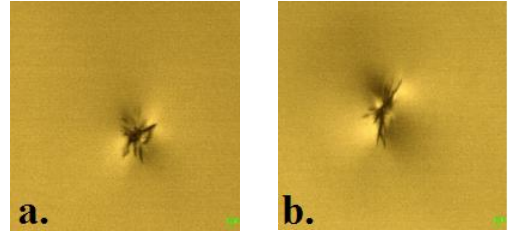

Fig. 3 Optical microscope images of rear surface damage at λ= 532nm; a. $t_p$= 500 ps, b. $t_p$= 8 ns. Magnified 6x

Figure 3 shows the damage morphology induced on the rear surface of the glass sample at 4mJ energy. The induced damage generates a streak inside the bulk of the sample instead of circular morphology because of the Rayleigh range of the focusing optics.

3.2 Damaged Surface Morphology

Surface morphologies induced by laser pulse at intensities above the damage threshold in BK7 glass are studied by using SEM. Figure 4a-h shows the SEM images of irradiated glass surface by picosecond (500 ps, 532 nm, single shot) and nanosecond (8ns, 532 nm, single shot) laser pulses.When the incident laserenergy density exceeds the critical value ($E_{pulse} > damage\ threshold$), the local temperature increases sharply and reaches the boiling point of the material. Due to this, generated tensile stress exceeds the dynamic strength of the material which causes the micro-cracks. The continuous merging of these micro cracks leads to ejection of one or several fragments from the sample. The ejected molten material from the interaction zone splashes out on the surface of glass around the irradiation region. These traces of spalled material and ejected fragments can easily be seen in the below figures 4a-h. The results indicate that during the interaction process glass was melted at the interaction zone. When the incident laser pulses are stopped, a slow local solidification of melt occur[17]. This solidification process of melt is slow because of low thermal conductivity of glass.The pulse energy required to achieve ablation in nanosecond pulse is higher than the picosecond pulse.

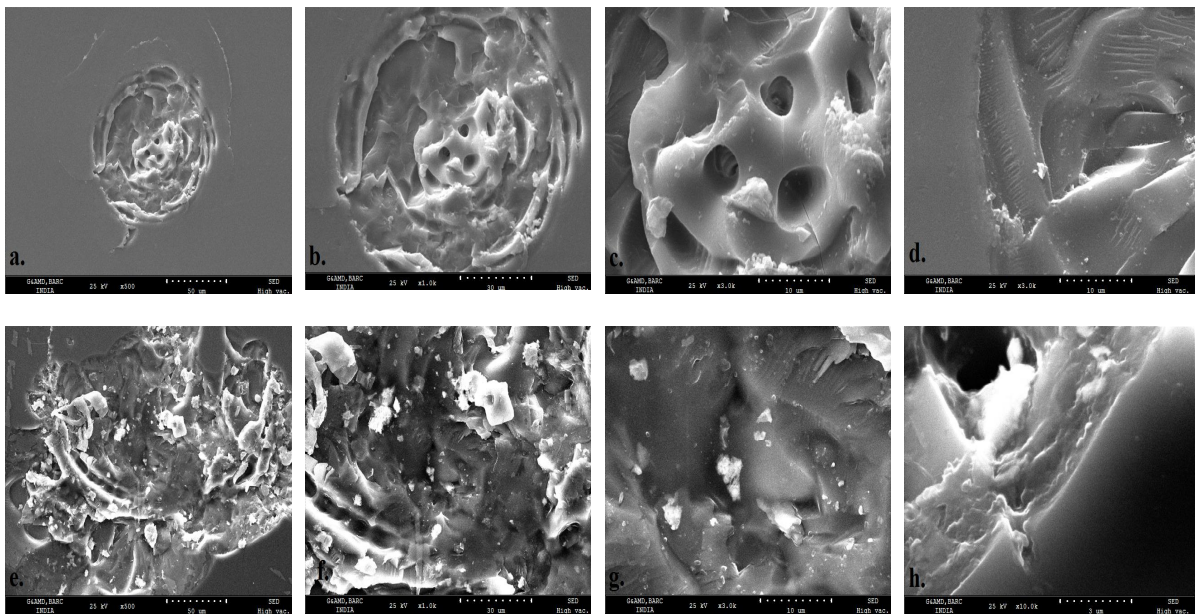

Fig.4 SEM images of irradiated glass target at intensity $3 \times 10^{10} W/cm^2$ (a-d) with $t_p$= 500 ps at magnifications (a) 500x (b) 1Kx (c) 3Kx and (d) 10Kx; (e-h) with $t_p$= 8 ns at magnifications (e) 500x (f) 1Kx (g) 3Kx (h) 10Kx

In the first image, figure 4a we can easily see the heat and mechanical stress affected zonearound the central crater. These cracks and heat affected zones are generated mainly due to the distribution of thermal heat and stress on the laser focused area at the time of plasma formation. With long pulses there is enough time for the thermal waves to propagate into the target and create relatively a large layer of melted material. On the other hand, in short pulse regime the lattice temperature remains much less than the electron temperature, thus laser ablation in this case accompanied by the electron heat conduction. By comparing figures 4a and 4e we can see that in case of nanosecond pulses sputtering and shock wave induce damage has taken place instead of melting process. While with picosecond pulse melting processes are dominated and at intensities $I_p \geq 3 \times 10^{10} W/cm^2$, a melt area like water waves, on the glass surface can be seen. Figure 4d and 4h shows the edge of damage surface, which suggest short pulses are more efficient in creation of precise and small hole with a lower energy, compared to long pulses.Brand and Tam reported that spherical molten droplets were ejected when an optical component surface was irradiated by picosecond lasers from the interaction zone[16], [20], [21]. While the irradiated surface region from glass generally was removed without melting process when exposed to nanosecond laser pulse.

3.3Effect of incident Laser energy on damaged area

Figure 5a-d shows measured induced damage width with respect to incident laser energy. From the graph it can be seen that the damage width increases nonlinearly and get saturated against the incident laser energy. The variation in rear surface damage width is small in comparison to the front surface, which is expected due less contamination deposition probability inside the bulk of the sample and possible self-focusing effect.

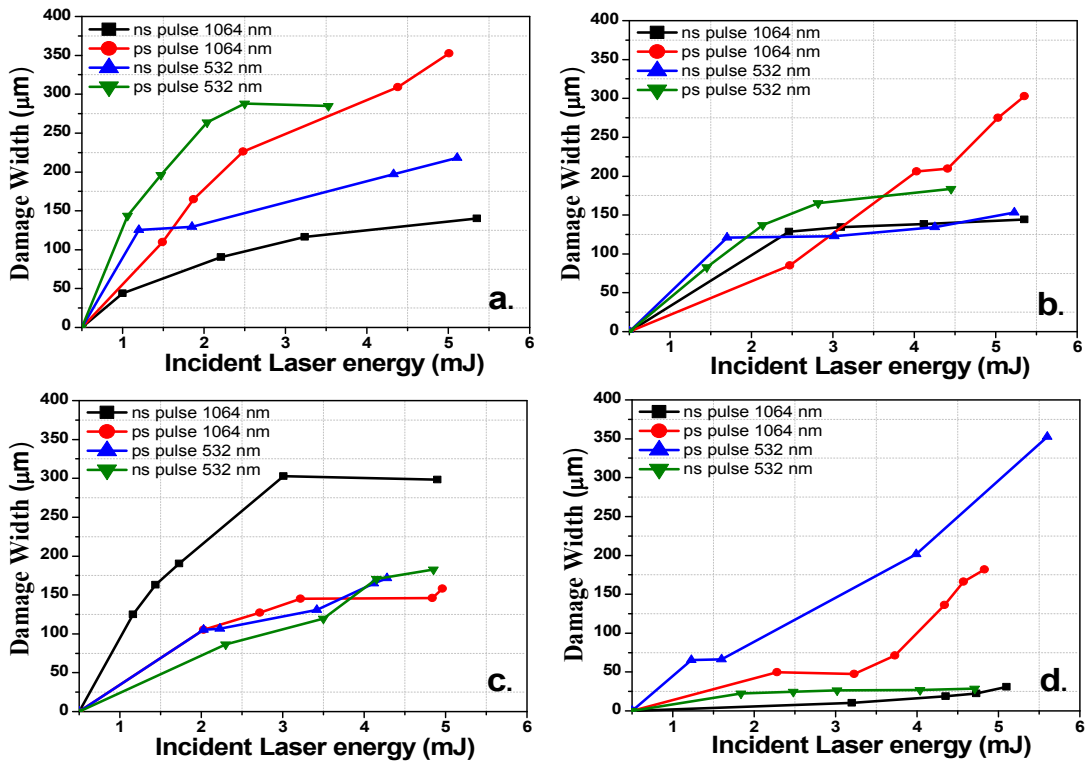

Fig.5 Effect of laser energy on damage width using laser parameters $t_p$= 500 ps, $t_p$= 8 ns; λ= 1064 nm, λ= 532 nm where;5a.(Front surface) and 5b.(Rear surface) showsdamage width variation with laser energy without $N_2$ flow on the surface;5c.(Front surface) and 5d.(Rear surface) shows damage width variation with laser energy with $N_2$ flow on the surface.

The magnitude of damaged area decreases significantly when a continuous gas flow is applied on the surface. Also it can be seen that the damage area depends on the sample impurity and is a statistical phenomenon. After a certain incident energy the increase rate in damage area is reduced due to the finite spot size of the incident beam and the damaged area tends toward the saturation.

3.4 Laser transmission through BK7 glass

The glass transmission is an important parameter for confinement geometry targets. Since intensity reaching the target and thus the resulting pressure depends upon the transmission from the plasma confining glass[13]. In this experiment, glass sample is irradiated by two different pulses (500 ps and 8 ns) and wavelengths (1064nm and 532nm). The transmitted laser pulse from the glass was collected by an energy meter. The percentage of transmitted laser energy as a function of incident laser energy is shown in figure 6a-d. As the laser intensity increases the percentage transmission through the glass decreases, showing more damage or absorption of laser light into the plasma. Also it can be seen that the laser transmission through back surface is more in comparison to the front surface. Since the front surface is more contaminated than the bulk, initiation of ionization occurs at lower energy. From the graphs it is clear that the laser energy absorption phenomenon happening here is completely nonlinear in nature and very much impurity dependent.

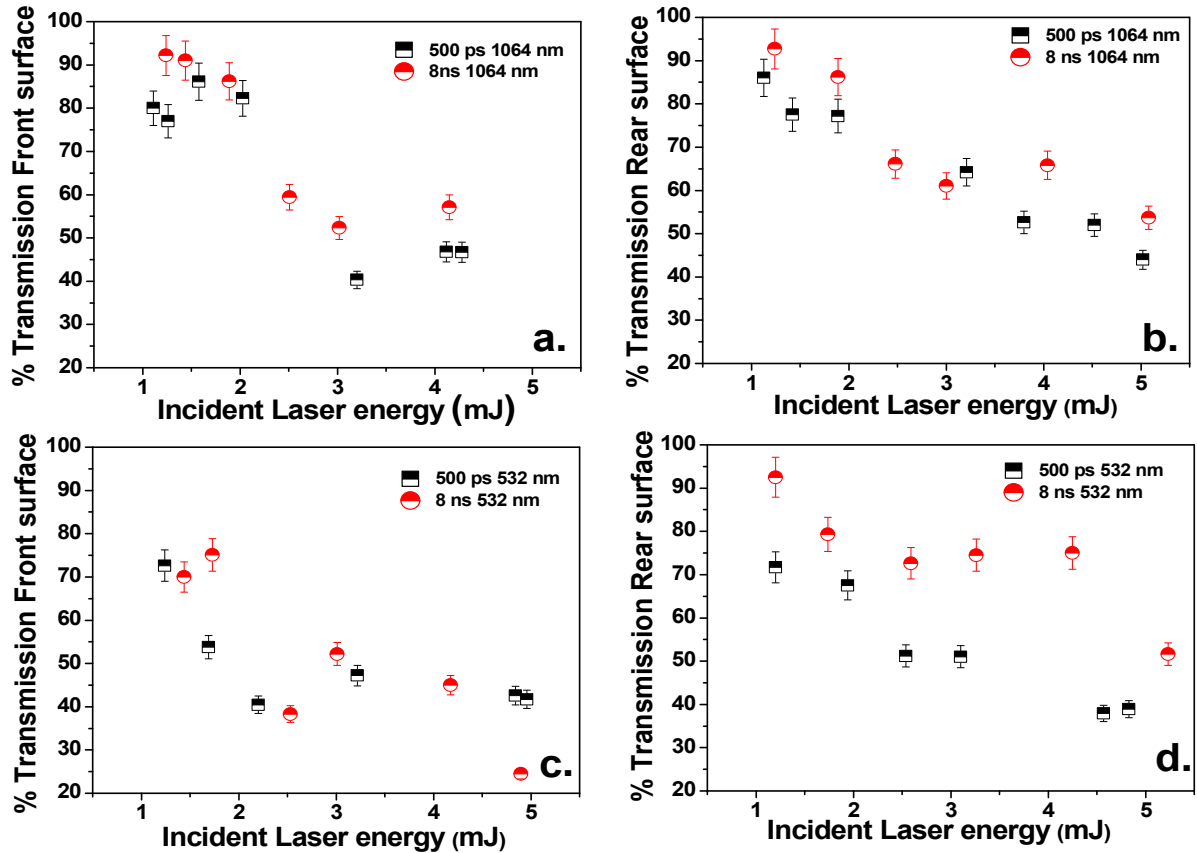

Fig.6 Percentage transmitted energy as a function of incident laser energy where; a. front surface % glass transmission at λ= 1064nm, $t_p$= 500ps and 8 ns; b. rear surface % glass transmission at λ= 1064nm, $t_p$= 500ps and 8 ns; c. front surface % glass transmission at λ= 532nm, $t_p$= 500ps and 8 ns; d. rear surface % glass transmission at λ= 532nm, $t_p$= 500ps and 8 ns.

3.5 Micro structural modifications in Borosilicate glass by laser irradiation using offline Raman spectroscopy

The Raman spectra of BK7 glass are studied to detect microstructural changes after laser irradiation; the results are shown in figure 7. Raman spectroscopy generally provides information about the changes in covalent network of the sample. For the undamaged portion of the sample, peaks are observed at 450, 518, 635, 810 and 1100 $cm^{-1}$. The peaks at 450 and 518 $cm^{-1}$ were assigned as bending modes of Si-O-Si linkages[22], [23]. Peak around 518 $cm^{-1}$ is related to the average inter-tetrahedral Si-O-Si angle[1], [22], [24]. The peak at 635 $cm^{-1}$ arises due to danburite ($B_2Si_2O_8^{2-}$) metastructural units and assigned to a single asymmetric mode. The peaks at 810 and 1100 $cm^{-1}$ were assigned to $SiO_4$ band and arises due to Si-O-Si linkages with two non-bridging oxygen atoms[1], [24]. Sharp peaks at 635 and 1100 $cm^{-1}$ indicates the intense vibration of $B_2Si_2O_8^{2-}$ and Si-O-Si groups.

The spectrum from damaged portion is very similar to undamaged portion. The vibrational band at 810cm$^{-1}$ shows a decrease in intensity, but does not vanishes completely by the laser irradiation which is due to their strong bond energy. For nanosecond laser damaged region the relative intensity of Raman peaks decreases significantly, which indicates a serious depolarization of vibrational bands caused by the irradiation. Compared to nanosecond laser irradiation the picoseconds laser irradiation damaged spectra have higher relative intensity, indicating that the vibration groups are more easily destroyed by nanosecond laser pulses. The unmodified spectra in damage sample suggests that the processes involved here in laser induced damages are an ionic transformation rather than a covalent conversion, since Raman spectroscopy is not much sensitive to charge transfer among ions.

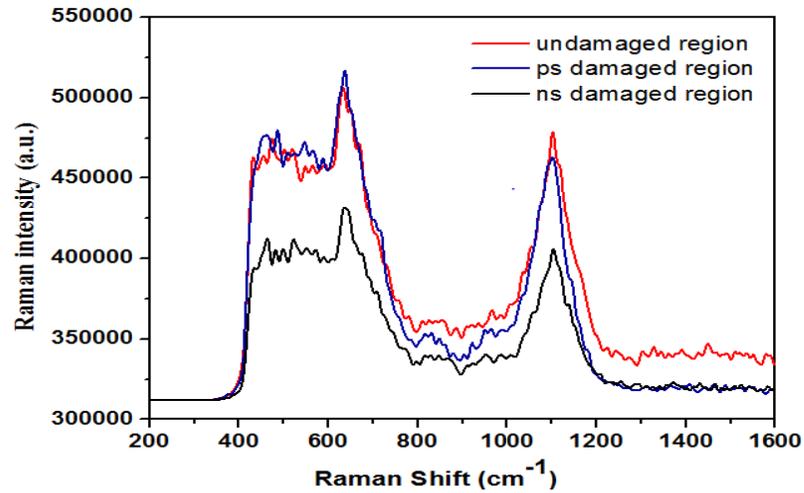

Fig. 7 Raman spectrum (after filtering) obtained from the damaged and undamaged region of BK7 glass at $t_p$= 500ps and 8ns

3.6 Optical Spectroscopy for plasma temperature and density calculation

The plasma formed during high power laser irradiation contains atoms and ions in different excited states, free electrons and radiation. The analysis of this plasma can be done through the measurement of plasma temperature ($T_e$) and free electron density ($n_e$). The plasma temperature describes the plasma state and the free electron density determines the thermodynamic equilibrium state of the plasma[25], [26]. The knowledge of the plasma temperature and density of the plasma species is important for understanding the atomic ionization and excitation processes occurring inside the plasma. The emitted spectrum is shown in figure 8.

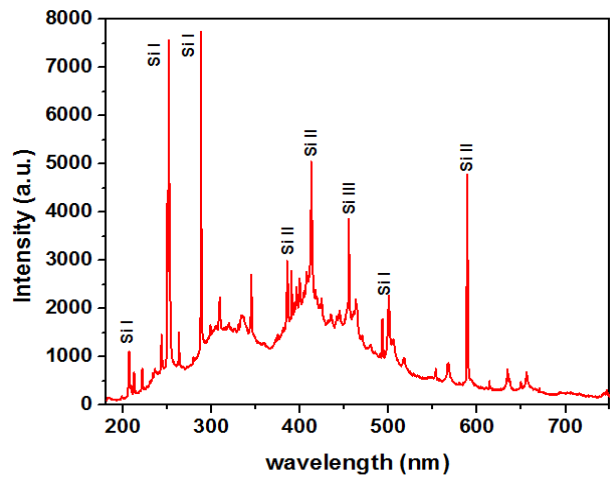

Fig. 8 Recorded emission spectra from plasma at λ= 1064 nm, $t_p$= 8 ns

### 3.6.1 Measurement of Plasma Temperature

The electron temperature measurement can be done by either of the following methods, namely; the relative intensity of two or more lines emitted from the same kind of species and same ionization stage or more general Boltzmann plot[25], [26].To avoid the corrections against the relative responses of the detector the wavelength separation of the line used must be very small and in order to get precise result the upper excited state energy separation should be large and the lines should be optically thini.e. absorption by plasma is negligible. For plasma in local thermodynamic equilibrium (LTE), the population density of atomic and ionic electronic states is described by Boltzmann distribution.The condition required for this assumption is that the radiative depopulation rates is negligible in comparison to collisional depopulation rate. For LTE plasma the temperature can becalculated from the slope of the lines defined by the below expression;

$$\ln\left(\frac{I_{ki}\lambda_{ki}}{g_k A_{ki}}\right) = C - E_k/k_B T \qquad (3)$$

Where $I, \lambda, A, E_k, g_k = 2J_k + 1$ are the spectral intensity, wavelength, transition probability, upper level energy and statistical weight of the upper state respectively, $C = \ln[hcN_k/4\pi Q(T)]$ [where Q (T) is the partition function and$N_k =$] and $T$ is the temperature.

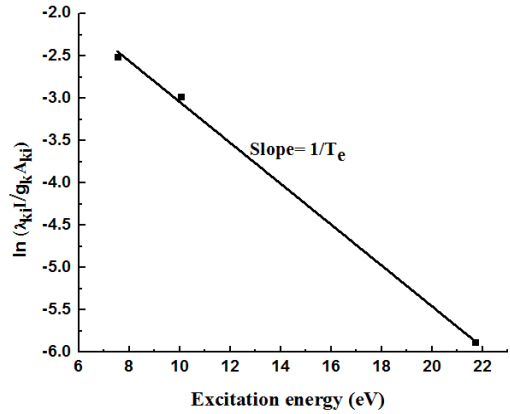

Fig.9 Saha Boltzmann plot using Silicon excitation lines for plasma temperature estimation

In LTE all temperatures are assumed to be equal, i.e. $T_e \approx T_i \approx T_{plasma}$. The spectral line wavelength, transition probabilities, statistical weight and energies of the upper levels are obtained from National Institute of Standard and Technology (NIST)[27]; shown in table 2. The plasma temperature estimated in this case is approximately 4 eV andthecorresponding Boltzmann plot is shown in figure 9.This estimated temperature is also verified by theoretical calculation using hydrodynamic simulations[28].The lines at 500 nm, 385nm and 455 nm are taken for the temperature calculations.

Table 2. Atomic parameters for the Si I, Si II and Si III used in temperature estimation

| Element | Wavelength $\lambda_{ki}$(nm) | Transition Probability $A_k$(s$^{-1}$) | Upper level Energy $E_k$(cm$^{-1}$) | Statistical Weight $g_k$ | Source Reference |
|---|---|---|---|---|---|
| Si I | 212.20 | 2.97×10$^8$ | 53362.24 | 7 | 27 |
| Si I | 500.17 | 2.8×10$^6$ | 60962.105 | 5 | 27 |
| Si II | 385.45 | 5.11×10$^6$ | 81251.32 | 4 | 27 |
| Si III | 455.12 | 1.26×10$^8$ | 175336.26 | 5 | 27 |

3.6.2 Measurement of free electron density

The most important quantity related to plasma density is the free electron density. If the relative abundance of species is known, ion densities can be calculated by using macroscopic neutrality condition from the measured free electron density. The electron number density emitted from plasma can be measured from the following methods; by themeasurement of optical reflectivity, by calculating the principal quantum number of the series limit, measurement of Stark profiles of optically thin emission spectral lines, measurementof absolute line densities and from the measurement of absolute emissivity of the continuum emission[25], [26].The Spectral lines width are always broaden and is a composition of finite resolution of the spectrometer used and intrinsic physical causes; the principal causes are Doppler and Stark broadening. If line widths are smallagainstradiative transition and perturbing level separations, Stark broadening is usually dominated by electron collisions. Since the Stark broadening of the spectral line depends on the electron number density, so electron density $n_e$ could be determined from the FWHM of the line. For a non H-like line the electrondensity can be calculated from the following relation[26], [29]

$$n_e(cm^{-3}) = \left(\frac{\Delta\lambda}{2\omega(\lambda,T_e)} N_r\right) \qquad (4)$$

Where, $\Delta\lambda$ is the FWHM of the corrected spectral line width, $\omega(\lambda, T_e)$ is the Stark broadening parameter can be found in the standard table[25], $N_r$ is the reference electron density. In the above expression the ionimpact broadening effect was neglected and assumed that lines used to evaluate electron number density should be optically thin.The calculated electron number density from the Lorentzian FWHM and atomic constant used to evaluate the electron densities is tabulated in Table 3. The calculated density for lines SiI and SiII are very close and is approximately $6.9\times10^{18}$ cm$^{-3}$.

Table 3. Spectral Parameters for Si I and Si II and calculated free electron density

| Element | Wavelength $\lambda_{ki}$ (nm) | Lorentzian FWHM $\Delta\lambda$ (nm) | Stark Broadening Parameter $\omega$ (nm)/$N_r$ | Reference Source | Calculated Electron Number Density $n_e$ (cm$^{-3}$) |
|---|---|---|---|---|---|
| Si I | 287.91 | 0.90 | 0.00064633/1×10$^{16}$ | 26 | 6.96×10$^{18}$ |
| Si II | 412.62 | 8.33 | 0.0606/1×10$^{17}$ | 26 | 6.87×10$^{18}$ |

4. Conclusions:

In this study, a detailed analysis of laser induced damage mechanism in BK7 glasswidely used for making confinement geometry targetshas been done. It is found that these induced damage geometries are functions of thelaser parameters such as laser wavelength, intensity, pulse duration, availability of free or seed electrons, material impurity etc. The samples display higher damage threshold energy when irradiated with long pulse in comparison to short laser pulse. Furthermore, due to reduced interaction time of pulse with the samples, the thermal losses are less in the picosecond regime as compared to nanosecond pulse.The estimated temperature and density for nanosecond produced plasma from the optical emission spectrum is around 4eV and $10^{18} cm^{-3}$, showing a subcritical plasma generation at the sample interaction interface. Lastly the laser modified and unmodified regions are studied with the help of Raman spectroscopy. The Raman results show that vibrational groups are more easily damaged by nanosecond laser pulses. The recorded Raman spectrahave similar nature for both the pulses and shows no big difference for damaged and undamaged regions, which suggest that the processes involved in the permanent transformation are ionic rather than bond or covalent conversion.

**Acknowledgement**

Authors wish to acknowledge Dr. S. M. Sharma, Director Physics group and Head HP&SRPD for his continuous support and encouragement received during the experiment. Authors are thankful to Dr. A. K. Sahu BARC, Mumbai, India for the Mini SEM SNE-3000M access.The authors are also thankful to Mr. C. G. Murali for his technical support during the experiment. Authors also want to acknowledge Mrs. P. Leshma for her valuable suggestions. SC thanks to Department of Atomic Energy, India for providing fellowship for Doctoral studies at HBNI, Mumbai, India.

**References**

[1] P. McMillan, "Structural studies of silicate glasses and melts-applications and limitations of Raman spectroscopy," *Am. Mineral.*, vol. 69, pp. 622–644, 1984.
[2] N. Bloembergen, "Laser-induced electric breakdown in solids," *IEEE J. Quantum Electron.*, vol. 10, no. 3, pp. 375–386, Mar. 1974.
[3] C. Carr, H. Radousky, and S. Demos, "Wavelength Dependence of Laser-Induced Damage: Determining the Damage Initiation Mechanisms," *Phys. Rev. Lett.*, vol. 91, no. 12, p. 127402, Sep. 2003.
[4] R. R. Gattass and E. Mazur, "Femtosecond laser micromachining in transparent materials," *Nat. Photonics*, vol. 2, no. 4, pp. 219–225, Apr. 2008.
[5] Z. Xia, D. Li, Y. Zhao, and Y. Wu, "New damage behavior induced by nanosecond laser pulses on the surface of silica films," *Opt. Laser Technol.*, vol. 46, no. 1, pp. 77–80, 2013.
[6] K. Sugioka and Y. Cheng, "Ultrafast lasers—reliable tools for advanced materials processing," *Light Sci. Appl.*, vol. 3, no. 4, p. e149, Apr. 2014.
[7] M. A. Meyers, F. Gregori, B. K. Kad, M. S. Schneider, D. H. Kalantar, B. A. Remington, G. Ravichandran, T. Boehly, and J. S. Wark, "Laser-induced shock compression of monocrystalline copper: characterization and analysis," *Acta Mater.*, vol. 51, no. 5, pp. 1211–1228, Mar. 2003.
[8] S. N. Luo, D. C. Swift, T. E. Tierney, D. L. Paisley, G. A. Kyrala, R. P. Johnson, A. A. Hauer, O. Tschauner, and P. D. Asimow, "Laser-induced shock waves in condensed matter: some techniques and applications," *High Press. Res.*, vol. 24, no. 4, pp. 409–422, Dec. 2004.
[9] J. Cheng, C. Liu, S. Shang, D. Liu, W. Perrie, G. Dearden, and K. Watkins, "A review of ultrafast laser materials micromachining," *Opt. Laser Technol.*, vol. 46, no. 1, pp. 88–102, Mar. 2013.
[10] A. Matsuda, T. Hongo, H. Nagao, Y. Igarashi, K. G. Nakamura, and K. Kondo, "Materials dynamics under nanosecond pulsed pressure loading," *Sci. Technol. Adv. Mater.*, vol. 5, no. 4, pp. 511–516, Jul. 2004.
[11] Z. A. Dreger and Y. M. Gupta, "High pressure Raman spectroscopy of single crystals of hexahydro-1,3,5-trinitro-1,3,5-triazine (RDX).," *J. Phys. Chem. B*, vol. 111, no. 15, pp. 3893–903, Apr. 2007.
[12] N. C. Anderholm, "Laser-Generated Stress Waves," *Appl. Phys. Lett.*, vol. 16, no. 3, p. 113, Oct. 1970.
[13] D. Devaux, R. Fabbro, L. Tollier, and E. Bartnicki, "Generation of shock waves by laser-induced plasma in confined geometry," *J. Appl. Phys.*, vol. 74, pp. 2268–2273, 1993.
[14] F. Cottet and M. Boustie, "Spallation studies in aluminum targets using shock waves induced by laser irradiation at various pulse durations," *J. Appl. Phys.*, vol. 66, no. 9, p. 4067, Nov. 1989.
[15] M. Boustie, T. De Rességuier, M. Hallouin, A. Migault, J. P. Romain, and D. Zagouri, "Some applications of laser-induced shocks on the dynamic behavior of materials," *Laser Part. Beams*, vol. 14, no. 02, p. 225, Mar. 2009.
[16] J. L. Brand and A. C. Tam, "Mechanism of picosecond ultraviolet laser sputtering of sapphire at 266 nm," *Appl. Phys. Lett.*, vol. 56, no. 10, p. 883, Mar. 1990.
[17] Y.-T. Chen, K.-J. Ma, A. A. Tseng, and P. H. Chen, "Projection ablation of glass-based single and arrayed microstructures using excimer laser," *Opt. Laser Technol.*, vol. 37, no. 4, pp. 271–280, Jun. 2005.
[18] Q. Zhang, F. Chen, N. Kioussis, S. Demos, and H. Radousky, "Ab initio study of the electronic and structural properties of the ferroelectric transition in $KH_2PO_4$," *Phys. Rev. B*, vol. 65, no. 2, p. 024108, Dec. 2001.
[19] Walter Koechner, Solid-State Laser Engineering .
[20] A. C. Tam, J. L. Brand, D. C. Cheng, and W. Zapka, "Picosecond laser sputtering of sapphire at 266 nm," *Appl. Phys. Lett.*, vol. 55, no. 20, p. 2045, Nov. 1989.
[21] M. R. Kasaai, V. Kacham, F. Theberge, and S. L. Chin, "The interaction of femtosecond and nanosecond laser pulses with the surface of glass," *J. Non. Cryst. Solids*, vol. 319, no. 1–2, pp. 129–135, May 2003.
[22] D. Manara, A. Grandjean, and D. R. Neuville, "Structure of borosilicate glasses and melts: A revision of the Yun, Bray and Dell model," *J. Non. Cryst. Solids*, vol. 355, no. 50–51, pp. 2528–2531, Dec. 2009.
[23] D. J. Little, M. Ams, S. Gross, P. Dekker, C. T. Miese, A. Fuerbach, and M. J. Withford, "Structural changes in BK7 glass upon exposure to femtosecond laser pulses," *J. Raman Spectrosc.*, vol. 42, no. 4, pp. 715–718, Apr. 2011.

[24] Q. Liu, B. F. Johnston, S. Gross, M. J. Withford, and M. J. Steel, "A parametric study of laser induced-effects in terbium-doped borosilicate glasses: prospects for compact magneto-optic devices," *Opt. Mater. Express*, vol. 3, no. 12, p. 2096, 2013.
[25] H. R. Griem, "Plasma Spectroscopy," vol. 35, pp. 34–130, 1966.
[26] A. M. El Sherbini, "Measurement of Plasma Parameters in Laser-Induced Breakdown Spectroscopy Using Si-Lines," *World J. Nano Sci. Eng.*, vol. 02, no. 04, pp. 206–212, Dec. 2012.
[27] N. US Department of Commerce, "NIST Atomic Spectra Database."
[28] C.D. Sijoy (Persional discussion).
[29] "N. Konjevic, A. Lesage, J. R. Fuhr and W. L. Wiese, 'Experimental Stark Widths and Shifts for Spectral Lines of Neutral and Ionized Atoms,' Journal of Physical and Chemical Reference Data, Vol. 31, No. 3, 2003.